\documentclass[iop]{emulateapj}

\def\ltsima{$\; \buildrel < \over \sim \;$}
\def\simlt{\lower.5ex\hbox{\ltsima}}
\def\gtsima{$\; \buildrel > \over \sim \;$}
\def\simgt{\lower.5ex\hbox{\gtsima}}
\def\kpc{{\rm\,kpc}}

\def\pc{{\rm\,pc}}

\def\deg{^\circ}
\def\s{\ifmmode \widetilde \else \~\fi}
\def\={\overline}

\def\spose#1{\hbox to 0pt{#1\hss}}

\def\lta{\mathrel{\spose{\lower 3pt\hbox{$\mathchar"218$}}
     \raise 2.0pt\hbox{$\mathchar"13C$}}}
\def\gta{\mathrel{\spose{\lower 3pt\hbox{$\mathchar"218$}}
     \raise 2.0pt\hbox{$\mathchar"13E$}}}
\def\Dt{\spose{\raise 1.5ex\hbox{\hskip3pt$\mathchar"201$}}}    % upper case
\def\dt{\spose{\raise 1.0ex\hbox{\hskip2pt$\mathchar"201$}}}    % lower case

\def\dotsfill{\leaders\hbox to 1em{\hss.\hss}\hfill}

\def\Gyr{{\rm\,Gyr}}
\def\FeH{{\rm[Fe/H]}}

\shorttitle{Three new Milky Way satellites in PS1}
\shortauthors{B. P. M. Laevens et al.}

\begin{document}

%% LaTeX will automatically break titles if they run longer than
%% one line. However, you may use \\ to force a line break if
%% you desire.

\title{Sagittarius~II, Draco~II and Laevens~3: three new Milky Way satellites discovered in the Pan-STARRS 1 3$\pi$ Survey}

%% Use \author, \affil, and the \and command to format
%% author and affiliation information.
%% Note that \email has replaced the old \authoremail command
%% from AASTeX v4.0. You can use \email to mark an email address
%% anywhere in the paper, not just in the front matter.
%% As in the title, use \\ to force line breaks.

\author{Benjamin P. M. Laevens$^{1,2}$, Nicolas F. Martin$^{1,2}$, Edouard J. Bernard$^{3}$, Edward F. Schlafly$^{2}$, Branimir Sesar$^{2}$, Hans-Walter Rix$^{2}$, Eric F. Bell$^{4}$, Annette M. N. Ferguson$^{3}$, Colin T. Slater$^{4}$, William E. Sweeney$^{5}$, Rosemary F. G. Wyse$^{6}$,  Avon P. Huxor$^{7}$, William S. Burgett$^9$, Kenneth C. Chambers$^{5}$, Peter W. Draper$^{8}$, Klaus A. Hodapp$^{5}$, Nicholas Kaiser$^{5}$, Eugene A. Magnier$^{5}$, Nigel Metcalfe$^{8}$, John L. Tonry$^5$, Richard J. Wainscoat$^{5}$, Christopher Waters$^{5}$}

\email{benjamin.laevens@astro.unistra.fr}

\altaffiltext{1}{Observatoire astronomique de Strasbourg, Universit\'e de Strasbourg, CNRS, UMR 7550, 11 rue de l'Universit\'e, F-67000 Strasbourg, France}
\altaffiltext{2}{Max-Planck-Institut f\"ur Astronomie, K\"onigstuhl 17, D-69117 Heidelberg, Germany}
\altaffiltext{3}{Institute for Astronomy, University of Edinburgh, Royal Observatory, Blackford Hill, Edinburgh EH9 3HJ, UK}
\altaffiltext{4}{Department of Astronomy, University of Michigan, 500 Church St., Ann Arbor, MI 48109, USA}
\altaffiltext{5}{Institute for Astronomy, University of Hawaii at Manoa, Honolulu, HI 96822, USA}
\altaffiltext{6}{Department of Physics and Astronomy, Johns Hopkins University, 3400 N. Charles St, Baltimore, MD 21218, USA}
\altaffiltext{7}{Astronomisches Rechen-Institut, Zentrum f\"ur Astronomie der Universit\"at Heidelberg, M\"onchhofstra§e 12 - 14, D-69120 Heidelberg, Germany}
\altaffiltext{8}{Department of Physics, Durham University, South Road, Durham DH1 3LE, UK}
\altaffiltext{9}{GMTO Corporation, 251 S. Lake Ave, Suite 300, Pasadena, CA  91101, USA}

\begin{abstract}
We present the discovery of three new Milky Way satellites from our search for compact stellar overdensities in the photometric catalog of the Panoramic Survey Telescope and Rapid Response System 1 (Pan-STARRS 1, or PS1) 3$\pi$ survey. The first satellite, Laevens~3, is located at a heliocentric distance of $d=67\pm3\kpc$. With a total magnitude of $M_V=-4.4\pm0.3$ and a half-light radius $r_h=7\pm2\pc$, its properties resemble those of outer halo globular clusters. The second system, Draco~II/Laevens~4 (Dra~II), is a closer and fainter satellite ($d \sim20\kpc$, $M_V =-2.9\pm0.8$), whose uncertain size ($r_h = 19^{+8}_{-6}\pc$) renders its classification difficult without kinematic information; it could either be a faint and extended globular cluster or a faint and compact dwarf galaxy. The third satellite, Sagittarius~II/Laevens~5 (Sgr~II), has an ambiguous nature as it is either the most compact dwarf galaxy or the most extended globular cluster in its luminosity range ($r_h = 37^{+9}_{-8}\pc$ and $M_V=-5.2\pm0.4$). At a heliocentric distance of $67\pm5\kpc$, this satellite lies intriguingly close to the expected location of the trailing arm of the Sagittarius stellar stream behind the Sagittarius dwarf spheroidal galaxy (Sgr dSph). If confirmed through spectroscopic follow up, this connection would locate this part of the trailing arm of the Sagittarius stellar stream that has so far gone undetected. It would further suggest that Sgr~II was brought into the Milky Way halo as a satellite of the Sgr dSph.
\end{abstract}


\section{Introduction}
Two decades ago the prevalent view of the Milky Way (MW) as an isolated system was radically changed by the discovery of a tidally disrupting dwarf galaxy, embedded in a stream in the constellation of Sagittarius \citep{ibata94}, highlighting the underrated importance of Milky Way-satellite interactions. With $\Lambda$CDM models predicting a whole new population of faint satellite dwarf galaxies (DGs) orbiting the MW \citep[e.g.][]{bullock00, bullock01}, the new challenge was to find these, until then, elusive objects. At the turn of the century, the advent of large CCD surveys such as the Sloan Digital Sky Survey (SDSS; \citealt{york00}) uncovered some 16 dwarf galaxies among the faintest ever found \citep[e.g.][]{willman05a, zucker06a, belokurov07a,walsh07}. Though revolutionizing our view of the satellite galaxies, just a handful of new globular clusters (GCs) were found, faint and nearby \citep{koposov07, belokurov10, munoz12b, balbinot13}. In addition, the SDSS enabled the discovery of several tidal streams (e.g. \citealt{belokurov06b, grillmair06b}), further illustrating the central role of satellite and cluster disruption in building up the MW's halo. 

With the second generation of surveys emerging such as the Dark Energy Survey (DES, \citealt{des05}), PS1 (Chambers et al. in preparation), later data releases of the SDSS, the Survey of the MAgellanic Survey History (SMASH, Nidever et al. in preparation), and VST Atlas have seen the number of known MW likely dwarf galaxies expand further from $\sim$25 to $\sim$35. These once elusive systems appear to be more common as deeper, but also wider coverage, data are gathered \citep{laevens15, bechtol15, koposov15, martin15, kim15a, kim15b}. In parallel, these systematic surveys also revealed a smaller number of faint new GCs \citep[e.g.][]{laevens14, kim15b}. The increase in the number of MW satellites led to the blurring of the traditional distinction between small, baryon-dominated GCs and more extended, dark-matter dominated DGs. Taking the photometric properties of these new satellites at face value shows that they straggle the DG and GC boundary in the size--luminosity plane, in the so-called ``valley of ambiguity" \citep{gilmore07}. Though follow-up observations have implied velocity dispersions higher than that expected from the stellar content for most of the new satellites \citep{martin07a,simon07,willman11,simon11,kirby13}, those measurements suffer from small number statistics and the unknown effect of binary stars on the kinematics of these small systems \citep{mcconnachie10}.

The recent discoveries of such faint candidate DGs out to $\sim70\kpc$ within DES confirm that they are in fact common and that they could indeed correspond to the large population of faint dark-matter dominated systems expected to inhabit the MW halo \citep[e.g.][]{tollerud08,bullock10}. These new satellites, located close to the Magellanic Clouds \citep{bechtol15,kirby15,koposov15}, have emphasized the tendency of these faint stellar systems to be brought into the MW surroundings in groups. Moreover, apparently isolated systems often share a proximity with stellar streams \citep[e.g.][]{belokurov08,deason14,laevens15,martin15}.

Over the last three years, our group has focused on the search for compact stellar systems in PS1, which has so far revealed two new MW satellites: a likely GC, Laevens~1/Crater \citep{laevens14, belokurov14b}, as well as a very faint satellite Triangulum~II \citep{laevens15}, whose nature has not yet been confirmed by spectroscopy. In this paper, we present the discovery of three new MW satellites discovered from the latest PS1 photometric catalog: a faint GC, Laevens~3 (Lae~3); a faint satellite, Draco~II/Laevens~4 (Dra~II), whose uncertain properties make its nature ambiguous; and another ambiguous system, Sagittarius~II/Laevens~5 (Sgr~II)\footnote{We assign double names for these last two systems, pending spectroscopic confirmation as to the nature of these stellar systems. For convenience and clarity, throughout the remainder of the paper, we refer to these satellites by their constellation name} \footnote{We assign Roman numeral II to this system and refer to the Sagittarius dwarf spheroidal galaxy discovered by \citet{ibata94} as Sgr dSph.}. This paper is structured in the following way: in section~2 we describe the PS1 survey and briefly outline the method which led to the discovery of the three satellites. In section~3 we discuss the properties of Lae~3, Dra~II, and Sgr~II, concluding and discussing the implications of the discoveries in section~4.

In this paper, all magnitudes are dereddened using the \citet{schlegel98} maps, adopting the extinction coefficients of \citet{schlafly11}. A heliocentric distance of $8\kpc$ to the Galactic center is assumed.

\section{The $3\pi$ PS1 Survey and discovery}
The PS1 survey (K. Chambers et al. in preparation) observed the whole sky visible from Hawaii ($\delta>-30\deg$), providing an unparalleled panoptic view of the MW and the Local Group. Throughout the 3.5 years of the $3\pi$ survey, the 1.8 m PS1 telescope, equipped with its 1.4-gigapixel camera capable of observing a 3.3-degree field of view, collected up to four exposures per year in five different optical filters: ($g_\mathrm{P1}r_\mathrm{P1}i_\mathrm{P1}z_\mathrm{P1}y_\mathrm{P1}$; \citealt{tonry12}). Once the individual frames have been taken at Haleakala and downloaded from the summit, the photometry is generated through the Image Processing Pipeline  \citep{magnier06,magnier07,magnier08}. 

The internal 3$\pi$ stacked catalogs were released in three processing versions (PV), with each consecutive version corresponding to a higher number of individual exposures and improved photometry. The three stellar systems described in this paper were found using the intermediate PV2 catalog and supplemented with the upcoming PV3 photometry for their analysis, when beneficial. Although there are many small differences between the two processing versions, their most interesting features for our study are that the PV2 psf photometry is performed on the stacked images, whereas for PV3, the stacks are only used to locate sources before performing the photometry on each individual sub-exposure, with its appropriate psf. As a consequence, the PV3 photometry is more accurate, but the PV2 star/galaxy separation is more reliable. The depths of the bands of PV2, enabling the discoveries, are comparable to the SDSS for the $g_\mathrm{P1}$ band (23.0) and reach $\sim0.5$/$\sim1.0$~magnitude deeper for $r_\mathrm{P1}$ (22.8) and $i_\mathrm{P1}$ (22.5; \citealt{metcalfe13}).

With large CCD surveys, automated search algorithms were developed to perform fast and efficient searches of these massive data sets for the small stellar overdensities that betray the presence of faint MW satellites. These techniques, originally implemented on the SDSS data \citep{koposov08,walsh09} have proven very successful. Inspired by this, we have developed our own similar convolution technique (Laevens et al. in preparation), adapted to the intricacies of the PS1 survey. The technique consists in isolating typical old, metal-poor DG or GC stars using de-reddened color-magnitude information $[(r-i)_0,i_0]$. For a chosen distance, masks in color-magnitude space are determined based on a set of old and metal-poor isochrones. The distribution of sources thereby extracted from the PS1 stellar catalog is convolved with two different window functions or Gaussian spatial filters \citep{koposov08}. The first Gaussian is tailored to the typical dispersion size of DGs or GCs ($2'$, $4'$, or $8'$), whereas the second one accounts for the slowly-varying contamination on far larger scales ($28'$ and $56'$). Subtracting the map produced from convolving the data with the larger Gaussian from that obtained with the smaller Gaussian results in maps of the PS1 sky tracking over- and under-densities once we further account for the small spatial inhomogeneities present in the survey. After cycling through different distances and the aforementioned sizes, we convert and combine all the density maps into statistical significance maps, allowing for a closer inspection of highly significant detections that do not cross-match with known astronomical objects (Local Group satellites, background galaxies and their GC systems, or artifacts produced by bright foreground stars). We further weed out spurious detections, by checking that these over densities do not correspond to significant background galaxy overdensities \citep{koposov08}. Applied to PV1, this method already led to the discovery of the most distant MW globular cluster Laevens~1/Crater \citep{laevens14,belokurov14b}, as well as one of the faintest MW satellites, Triangulum~II \citep{laevens15}, whose nature is not yet known. Sgr~II, Dra~II, and Lae~3 were detected as 11.9, 7.4, and 6.5 $\sigma$ detections, comfortably above our $5\sigma$ threshold\footnote{For context, applying this technique leads to the recovery of the faint satellites Segue~1 and Bootes~I, originally discovered in the SDSS, with a significance of $7.3$ and $11.6 \sigma$, respectively.}. All three new satellites lie outside the SDSS footprint, which explains why they were not discovered before. Sgr~II and Lae~3 are located at fairly low Galactic latitude\footnote{In fact, Lae~3 is clearly visible on the DSS plates and could have been discovered before the PS1 era.} ($b\sim-20\deg$) and Dra~II is quite far north ($\delta\sim+65\deg$).

\begin{figure*}[htb]
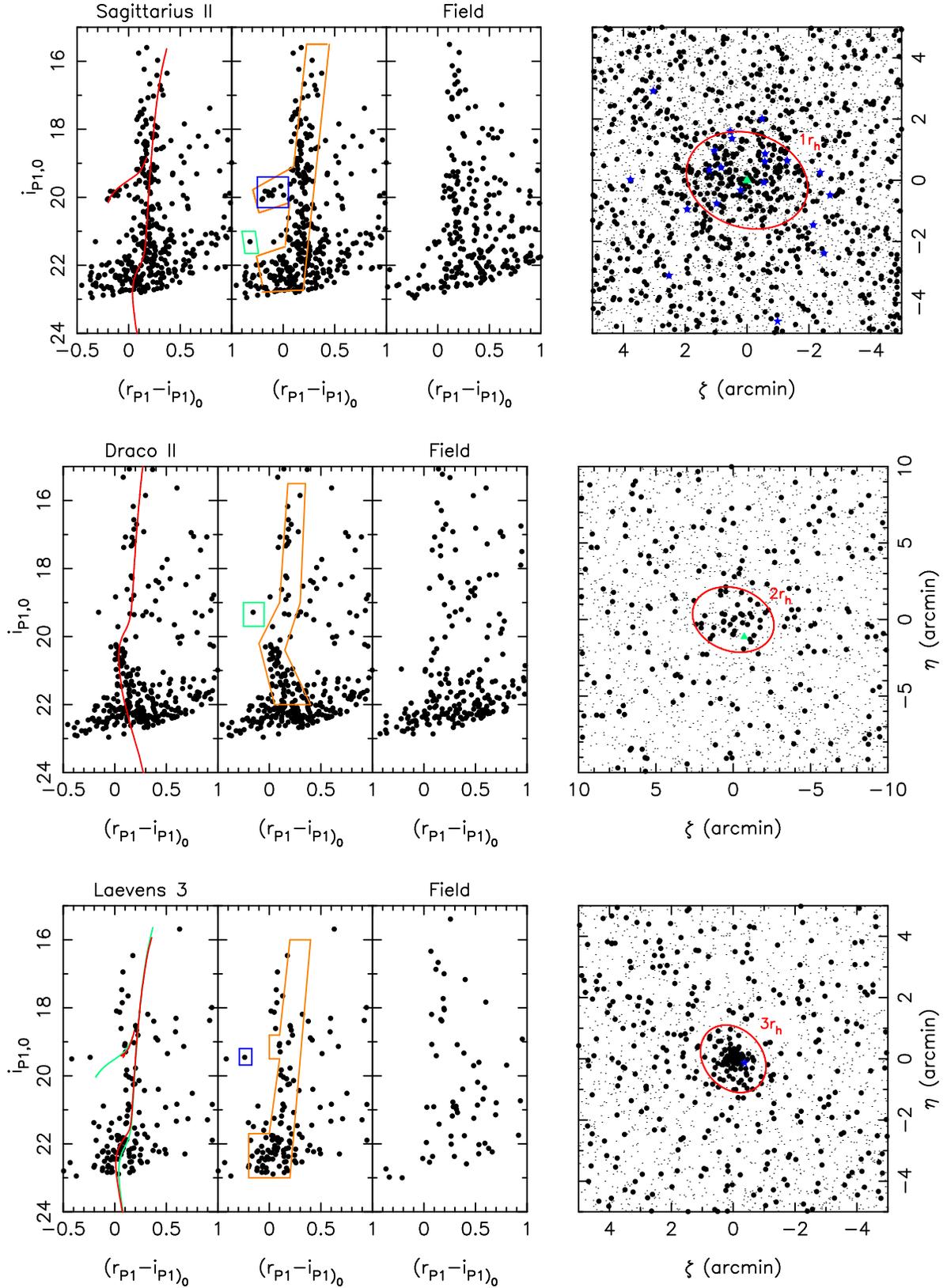

\centering
    \includegraphics[width=.38\textwidth,angle=270]{Figure_1a.ps}\\
  \vspace{0.5cm}   
    \includegraphics[width=.38\textwidth,angle=270]{Figure_1b.ps}\\
    \vspace{0.5cm}    
    \includegraphics[width=.38\textwidth,angle=270]{Figure_1c.ps}
  \caption{\label{CMDs}\emph{Left:} from top to bottom, CMD of stars within 1 half-light radius (Sgr~II), 2 half-light radii (Dra~II) or 3 half-light radii (Lae~3) with the favored isochrone: $12\Gyr$ and $\FeH\sim-2.2$ for Sgr~II and Dra~II, and $9\Gyr$ and $\FeH=-1.9$ (Lae~3). In the case of Lae~3, we also show the same old and metal--poor isochrone (green) as for Sgr~II and Dra~II, highlighting that with current photometry, it is difficult to distinguish between the two. \emph{Middle-left:} Same as left-most panel with the CMD selection box used to isolate the RGB, HB, MSTO, and/or MS stars of the satellites (orange), and an HB selection box (blue) for Sgr~II. The RR Lyra star for Lae~3 is highlighted in blue. Candidate blue straggler stars are identified for Sgr~II and Dra~II (green). \emph{Middle-right:} CMD of field regions for stars 15 arcmin West of the satellites of similar sizes to those used for the left-most panels. \emph{Right:} Spatial distribution of all stars around the three satellites (light black dots) and of stars selected with the orange CMD selection boxes in the middle-left panel (big black dots). For Sgr~II, HB stars corresponding to the blue HB selection box in the middle panel are represented by blue stars symbols. Finally, candidate blue straggler stars corresponding to the green box are displayed as green triangles.}
\end{figure*}

\section{Properties of the three stellar systems}
\subsection{Color-Magnitude Diagrams and Distances}
\subsubsection{Sagittarius~II}
The color-magnitude diagram (CMD) of stars within one half-light radius of Sgr~II ($2.0'$; see below for the structural parameters) is displayed in the top row of panels in Figure~\ref{CMDs}, next to the CMD of a field region of the same size. Since the features of Sgr~II are so obvious we rely here only on the more accurate PV3 photometry at the cost of a poorer star/galaxy separation at the faint end. Given the location near the MW bulge [$(\ell,b) = (18.9\deg,-22.9\deg)$] the field CMD is very populated, but the Sgr~II CMD features are nevertheless clearly defined with a red giant branch (RGB) visible between $[(r_\mathrm{P1}-i_\mathrm{P1})_0,i_\mathrm{P1,0}] \simeq [0.30,16.5] $ and [0.15,21.5], before its main sequence turnoff at $i_\mathrm{P1,0}<22.0$. The most obvious feature, however, is the horizontal branch (HB) of the system, clearly visible for $19.5<i_\mathrm{P1,0}<20.0$ and $(r_\mathrm{P1}-i_\mathrm{P1})_0<0.0$. When selected with the box overlaid in orange on the CMD of the second Sgr~II CMD panel of Figure~\ref{CMDs}, these stars correspond to a well-defined spatial overdensity (right-most panel). Isolating the HB stars (blue box) also highlights how clustered they are on the sky. We further highlight a single star that is bluer and brighter than the turn off and could potentially correspond to a blue straggler (green box and green triangle).

The presence of the reasonably well-populated HB, with 13 stars within 3 half-light radii, allows for a robust estimation of the distance to the satellite. Equation 7 in \citep{deason11} describes the relation between the absolute magnitude of these HB stars and their SDSS $g-r$ colors. Converting the PS1 magnitudes to the SDSS bands, for which the relation holds, reveals a median $g=19.60\pm 0.03$ and $M_{g}=0.47\pm0.04$  when we perform a Monte Carlo resampling of the stars' uncertainties. These lead to a distance-modulus of $19.13\pm0.15$, where an uncertainty of 0.1 was assumed on the \citet{deason14} relation. This translates into a Heliocentric distance of $67\pm5$kpc or a Galactocentric distance of $60\pm5$ kpc. Fixing the satellite at this distance modulus, we experiment with isochrones. Overlaid on the Sgr~II CMD of Figure~\ref{CMDs}, we also show the old and metal-poor isochrone from the \textsc{Parsec} library \citep[12\Gyr, $\FeH=-2.2$; ][]{bressan12} that provides the best qualitative fit to the CMD features at this distance.

The properties of Sgr~II are summarized in Table~1.

\subsubsection{Draco~II}
Draco~II is much closer and less luminous than Sgr~II, as can be seen in the CMD of stars within $2r_h$ of the satellite's center in the second row of panels in Figure~\ref{CMDs}. Here, since we need both depth and a good star/galaxy separation to clean the main sequence of Dra~II, we use the PV3 photometry combined with the superior PV2 star/galaxy flags. This has the consequence of removing some faint PV3 stars misidentified as galaxies but more optimally cleans the main sequence of the satellite. A field CMD is shown in the right-most CMD panel and helps identify the Dra~II features: a populated main sequence between $[(r_\mathrm{P1}-i_\mathrm{P1})_0,i_\mathrm{P1,0}] \simeq [0.0,20.2]$ and [0.2,22.0]. At brighter magnitudes, Dra~II shows no HB and no prominent RGB. However, a group of stars at [0.2,17.0] is compatible with being the system's sparsely sampled RGB. As for Sgr~II, isolating the stars in these CMD features (orange box in the central CMD panel) highlights the stellar overdensity in the spatial distribution shown in the right-most panel. As for Sgr~II, we identify a potential blue straggler in green.

Due to the absence of any HB star\footnote{This is not per se surprising as, for instance, the similarly faint system Willman~1 only contains two HB stars \citep{willman11}.}, we cannot reliably break the distance-age-metallicity degeneracy with the PS1 data alone. Consequently, we explored isochrones of different ages and metallicities, located at varying distance. The best fit is provided by the \textsc{Parsec} isochrone shown in Figure~\ref{CMDs}; it has an age of 12\Gyr~and $\FeH = -2.2$ and is located at a distance modulus of $16.9\pm0.3$ but we caution the reader on the reliability of this particular isochrone that needs to be confirmed from deeper data.

\begin{figure}
\begin{center}
\includegraphics[width=0.96\hsize]{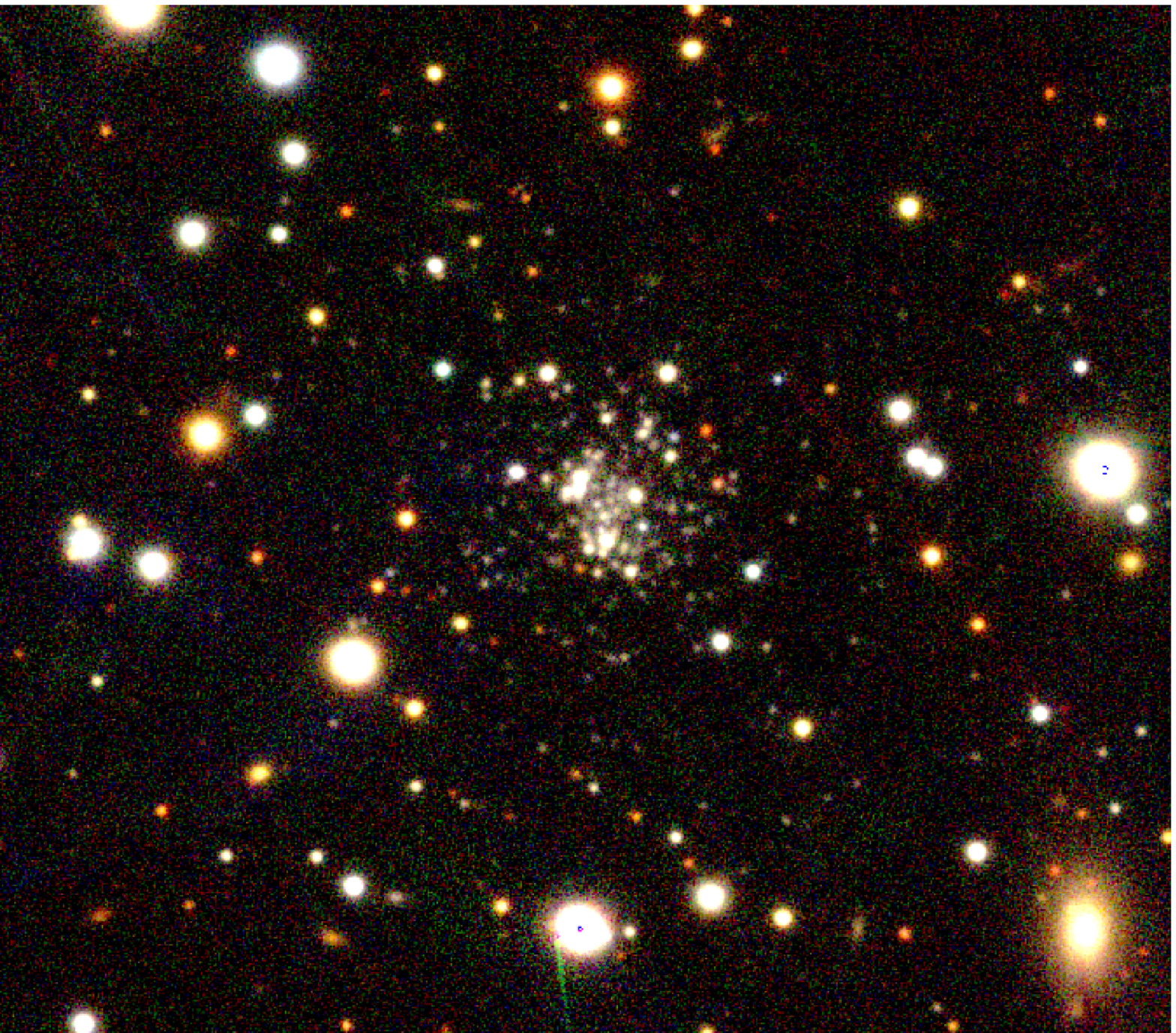}
\caption{\label{Postage_Stamp}: $g_\mathrm{P1}r_\mathrm{P1}i_\mathrm{P1}$ image of Laevens~3, built from the stacked PV3 PS1 images. The image is $2.5'\times2.5'$ and North is to the top and East is to the left.}
\end{center}
\end{figure}

\subsubsection{Laevens~3}
As can be seen in Figure~\ref{Postage_Stamp}, Lae~3 is a compact stellar system\footnote{Note that the PS1 postage stamp images show no clear stellar over density for Sgr~II and Dra~II, hence why we do not include them.}. As such, the automated PS1 pipeline fails to extract the photometric information of the central region that suffers from crowding. We therefore perform custom photometry of this sky cell using \textsc{daophot}, using the same method as \citep{laevens14}. The resulting CMD for stars within $3r_h$ of the stellar system's centroid is shown on the bottom row of panels in Figure~\ref{CMDs}. This CMD is still likely to suffer from crowding, yet it reveals features that are clearly not expected in the field population: the Lae~3 RGB between $[(r_\mathrm{P1}-i_\mathrm{P1})_0,i_\mathrm{P1,0}] \simeq [0.75,18.0]$ and [0.15,21.0], followed by the system's MSTO at fainter magnitudes. The stars between $[(r_\mathrm{P1}-i_\mathrm{P1})_0,i_\mathrm{P1,0}] \simeq [0.0,16.2] $ and [0.2,18.2] are foreground contaminants and are situated far from the satellite center, just under the $3r_h$ limit. As for the two other satellites, selecting these stars only (orange box in the middle CMD) highlight a clear stellar overdensity.

An investigation into the presence of RR Lyra stars in the PS1 temporal data (Hernitschek et al. 2015, in prep.; Sesar et al. 2015, in prep.) reveals one obvious candidate, 0.6 arcmins away from the center of the cluster (highlighted by the blue box in the middle CMD and represented by a blue star in the right-most panel).  Briefly, RR Lyrae stars are identified in PS1 data by providing average PS1 colors and various variability statistics to a trained Random Forest classifier \citep{richards11}. The resulting RR Lyrae sample is 80\% complete (up to 80 \kpc) and 90\% pure. The distances of PS1 RR Lyrae stars are uncertain at the 5\% level. The RR Lyra star in Lae~3 has also been observed more than 100 times by the Palomar Transient Factory (PTF; \citealt{law09, rau09}). The distance and period measured from PS1 data agree within $2\kpc$ and 5\% with those measured from the PTF data. The RR Lyra star is at $(m-M)_{0}=19.14\pm0.10$, or a distance of $67\pm3\kpc$. The $\sim$14 hour period of the star suggests a star with a metallicity range of $-1.9<\FeH<-1.6$. Fixing this distance, we once again experiment with various isochrones and conclude that the CMD feature are best tracks by an isochrone with a comparatively young age of 9\Gyr~and $\FeH\sim-1.9$, compatible with the properties of the RR Lyra star. The isochrone fit, fixed at that distance, tracks the main features of the satellite such as the RGB and the MSTO. Though two blue stars are also present at the same magnitude as the tentative red HB, the bluest of the two is a field variable star, incompatible for being a member of Lae~3.
We further compare the CMD features with the GC fiducial published by \citet{bernard14} in the PS1 photometric system. The fiducials of globular clusters NGC 1904, NGC 5897 and NGC 7089, with $-1.9<\FeH<-1.6$ provide a good fit to the CMD features and confirm our impression from the isochrones.

The derived properties are summarized in Table~1 for all three satellites.

\begin{table*}
\caption{\label{properties}Properties of Laevens~3, Draco~II and Sagittarius~II}
\begin{center}
\begin{tabular}{cccc}
\hline\hline
& Laevens~3 & Draco~II/Laevens~4 & Sagittarius~II/Laevens~5\\
\hline
  $\alpha$    (ICRS)                &    21:06:54.3             &    15:52:47.6             &    19:52:40.5             \\
  $\delta$    (ICRS)                &    +14:58:48              &    +64:33:55              &    $-$22:04:05            \\
  $\ell       (\deg)$            &    63.6                   &    98.3                   &    18.9                   \\
  $b             (\deg)$            &    $-21.2$                &    +42.9                  &    $-22.9$                \\
  Distance Modulus               &    $19.14\pm0.10$      &    $\sim16.9\pm0.3$   &    $19.13\pm0.15$      \\
  Heliocentric Distance (kpc)         &    $67\pm3$            &    $20\pm3$            &    $67\pm5$            \\
  Galactocentric Distance (kpc)         &    $64\pm3$            &    $22\pm3$            &    $60\pm5$            \\
  $M_{V}$                             &    $-4.4\pm0.3$        &    $-2.9\pm0.8$        &    $-5.2\pm0.4$        \\
  $L_V$                               &    $10^{3.7\pm0.1}$    &    $10^{3.1\pm0.4}$    &    $10^{4.0\pm0.1}$    \\
  $\FeH$                             &    $\sim-1.9$          &    $\sim-2.2$          &    $\sim-2.2$          \\
  Age            (Gyr)                 &    $\sim9$             &    $\sim12$            &    $\sim12$            \\
  $E(B-V)^{a}$                        &    0.073                  &    0.016                  &    0.097                  \\
  Ellipticity                         &    $0.21\pm0.21$       &    $0.24^{+0.27}_{-0.24}$ &    $0.23^{+0.17}_{-0.23}$ \\
  Position       angle (from N to E $\deg$) &    $40^{+16}_{-28}$       &    $70\pm28$           &    $72^{+28}_{-20}$       \\
  $r_{h}$        (arcmin)              &    $0.40^{+0.07}_{-0.11}$ &    $2.7^{+1.0}_{-0.8}$    &    $2.0^{+0.4}_{-0.3}$    \\
  $r_{h}$        (pc)                  &    $7\pm2$             &    $19^{+8}_{-6}$         &    $38^{+8}_{-7}$         \\
\hline
$^a$ from \citet{schlegel98} and \citet{schlafly11} &\\
\end{tabular}
\end{center}
\end{table*}

\begin{figure}
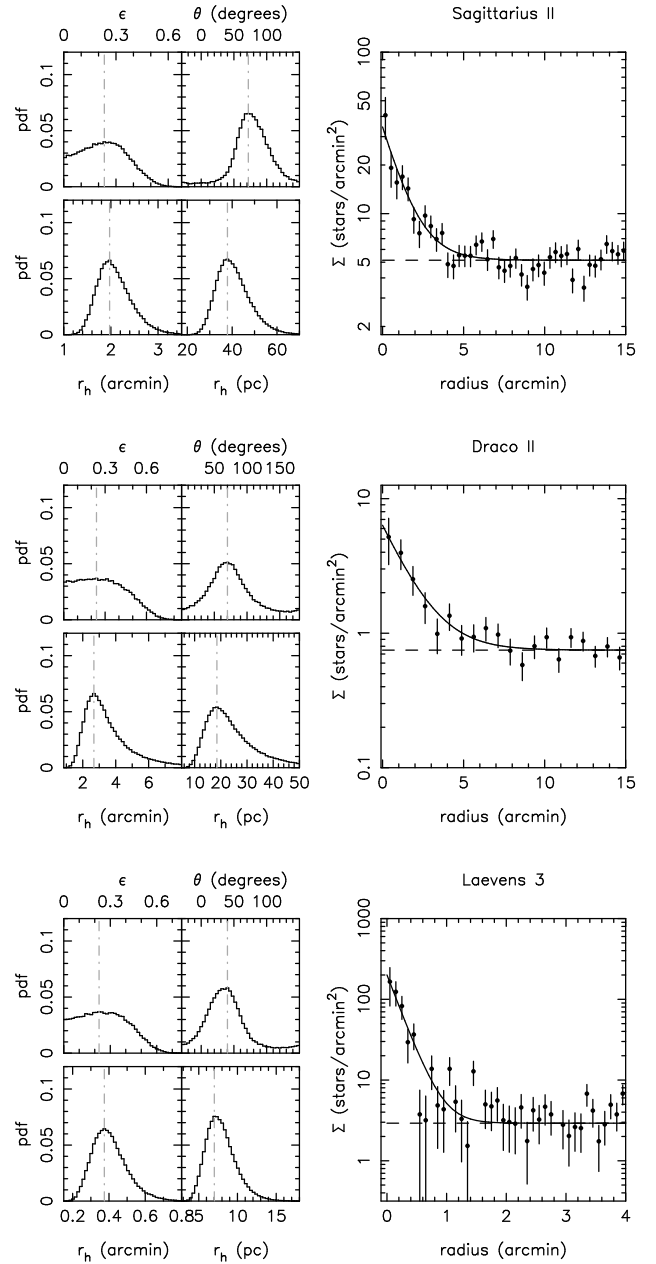

\includegraphics[width=.29\textwidth,angle=270]{Figure_3a.ps}
\vspace{0.5cm}\\
\includegraphics[width=.29\textwidth,angle=270]{Figure_3b.ps}
\vspace{0.5cm}\\
\includegraphics[width=.29\textwidth,angle=270]{Figure_3c.ps}
\caption{\label{MCMC} \emph{Left:} Probability distribution functions for the ellipticity ($\epsilon$), the position angle ($\theta$), and the angular and physical half-light radii, $r_{h}$, of Sgr~II (top), Dra~II (middle), and Lae~3 (bottom). \emph{Right:} Comparison between the favored radial distribution profile (full line) and the data, binned according to the preferred structural parameters (dots), selected as the mode of the PDFs (grey line in the left-hand panels). The error bars assume Poissonian uncertainties and the dashed line represents the field density.}
\end{figure} 

\subsection{Structural Parameters and Absolute Magnitudes}

The structural parameters of Sgr~II, Dra~II, and Lae~3 are derived using a version of the \citet{martin08b} likelihood technique updated to a full Markov Chain Monte Carlo framework (Martin et al. 2015, in prep.). Using the star's location in the vicinity of satellite, the algorithm calculates the posterior probability distribution function (PDF) of a family of exponential radial density profiles, allowing for flattening and a constant contamination from field stars. The parameters of the models are the centroid of the system, the ellipticity, $\epsilon$\footnote{The ellipticity, here, is defined as $\epsilon=1-b/a$ with $a$ and $b$ the major and minor axis scale lengths, respectively.}, the position angle, $\theta$ (defined as the angle of the major axis East from North), the half-light radius $r_{h}$\footnote{In fact, the algorithm constrains the half-density radius, but this is similar to the more common half-light radius if there is no mass segregation in the system.}, and finally the number of stars, $N^*$ within the chosen CMD selection box. We further determine the physical half-light radius from the angular one by randomly drawing distances from the distance modulus values.

The PDF for the ellipticity, position angle, as well as the angular and physical half-light radii may be seen in Figure~\ref{MCMC} for, from top to bottom, Sgr~II, Dra~II, and Lae~3. The three right-most panels of the figure compare the radial profile of a given satellite, binned following the favored centroid, ellipticity, and position angle to the favored exponential radial density profile; they display the good quality of the fit in all cases. All three systems are rather compact, with angular half-light radii of $2.0^{+0.4}_{-0.3}$, $2.7^{+1.0}_{-0.8}$, and $0.40^{+0.07}_{-0.11}$ arcmin for Sgr~II, Dra~II, and Lae~3, respectively. However, the different distances to these systems lead to different physical sizes: $38^{+8}_{-7}$, $19^{+8}_{-6}$, and $7\pm2\pc$. In all three cases, the systems appear mildly elliptical but the PDFs show that this parameter is poorly constrained from the current data. It should be noted that, in the case of Lae~3, the crowding at the center of the stellar system could lead to an underestimation of the compactness and luminosity of the system. However, the Lae~3 radial profile shows no sign of a central dip.

The absolute magnitude of the three stellar systems was determined using the same procedure as for Laevens~1 and Triangulum~II \citep{laevens14, laevens15}, as was also described for the first time in \citet{martin08b}. Using the favored isochrones and their associated luminosity functions for the three satellites, shifted to their favored distances, we build CMD pdfs after folding in the photometric uncertainties. Such CMDs are populated until the number of stars in the CMD selection box equals the favored number of stars $N^*$ as determined by the structural parameters\footnote{For this part of the analysis, we make sure to use a selection box that remains $\sim1$ magnitude brighter than the photometric depth so the data is close to being complete. In the case of Lae~3, the sparsely populated CMD prevents us from doing so as we need to use the full extent of the CMD to reach convergence in the structural parameter analysis. This is likely to slightly underestimate the luminosity of the cluster.}. The flux of these stars is summed up, yielding an absolute magnitude. In practice, this operation is repeated a hundred times with different drawings of the Markov chains, thus taking into account three sources of uncertainty: the distance modulus uncertainty, the uncertainty on the number $N^*$ of stars in the CMD selection box, and shot-noise uncertainty, originating from the random nature of populating the CMD. This procedure yields total magnitudes in the PS1 $r_\mathrm{P1}$ band, which we then convert to the more commonly used $V$-band magnitudes through a constant color offset ($V-r=0.2$) determined from the analysis of more populated, known, old and metal-poor MW satellites. This yield $M_{V}=-5.2\pm0.4$, $-2.9\pm0.8$, and $-4.4\pm0.3$ for Sgr~II, Draco~II, and Laevens~3, respectively. All three systems are rather faint, as expected from their sparsely populated CMDs.

\begin{figure*}[htb]
\centering
    \includegraphics[width=.5\textwidth,angle=270]{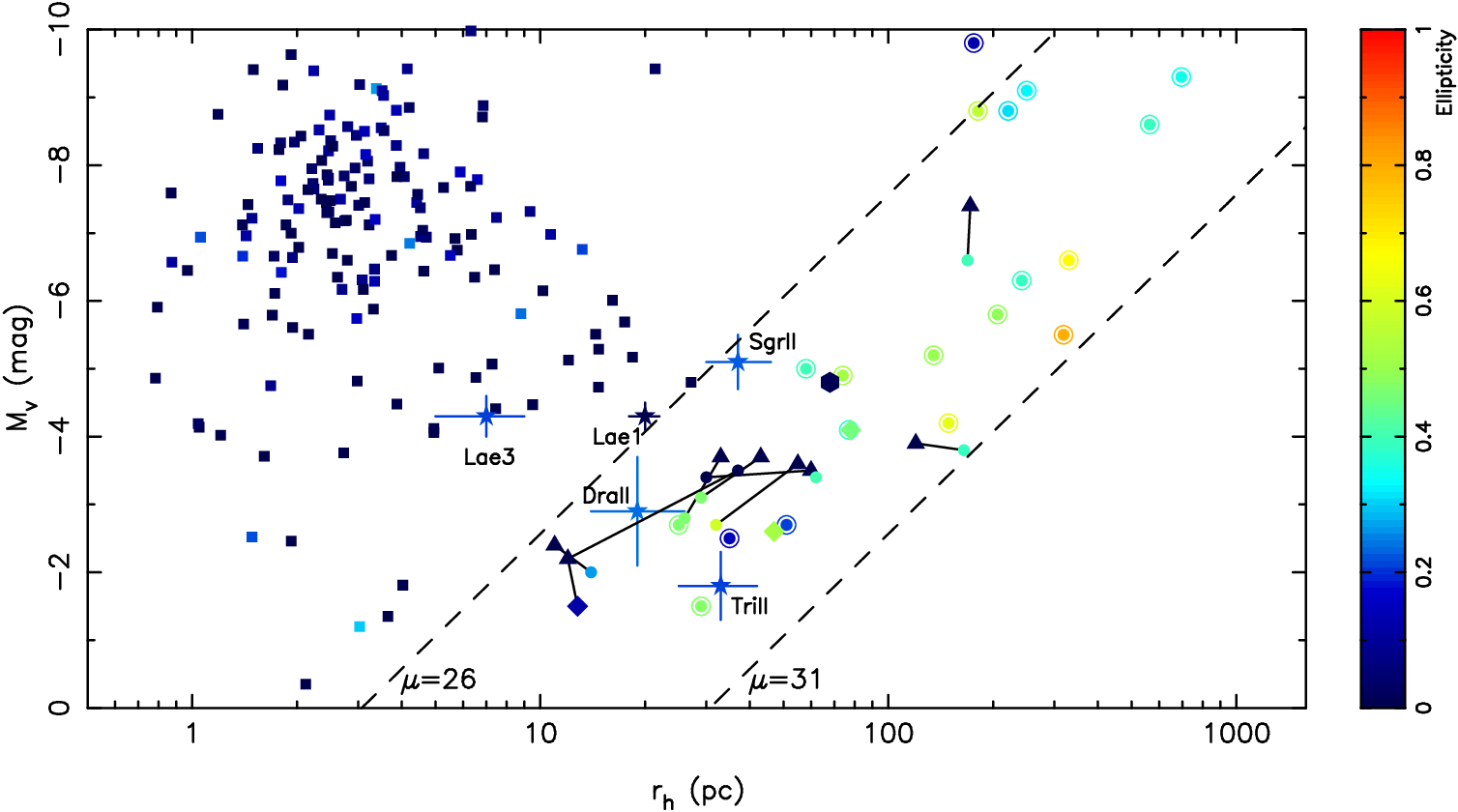}
    \includegraphics[width=.5\textwidth,angle=270]{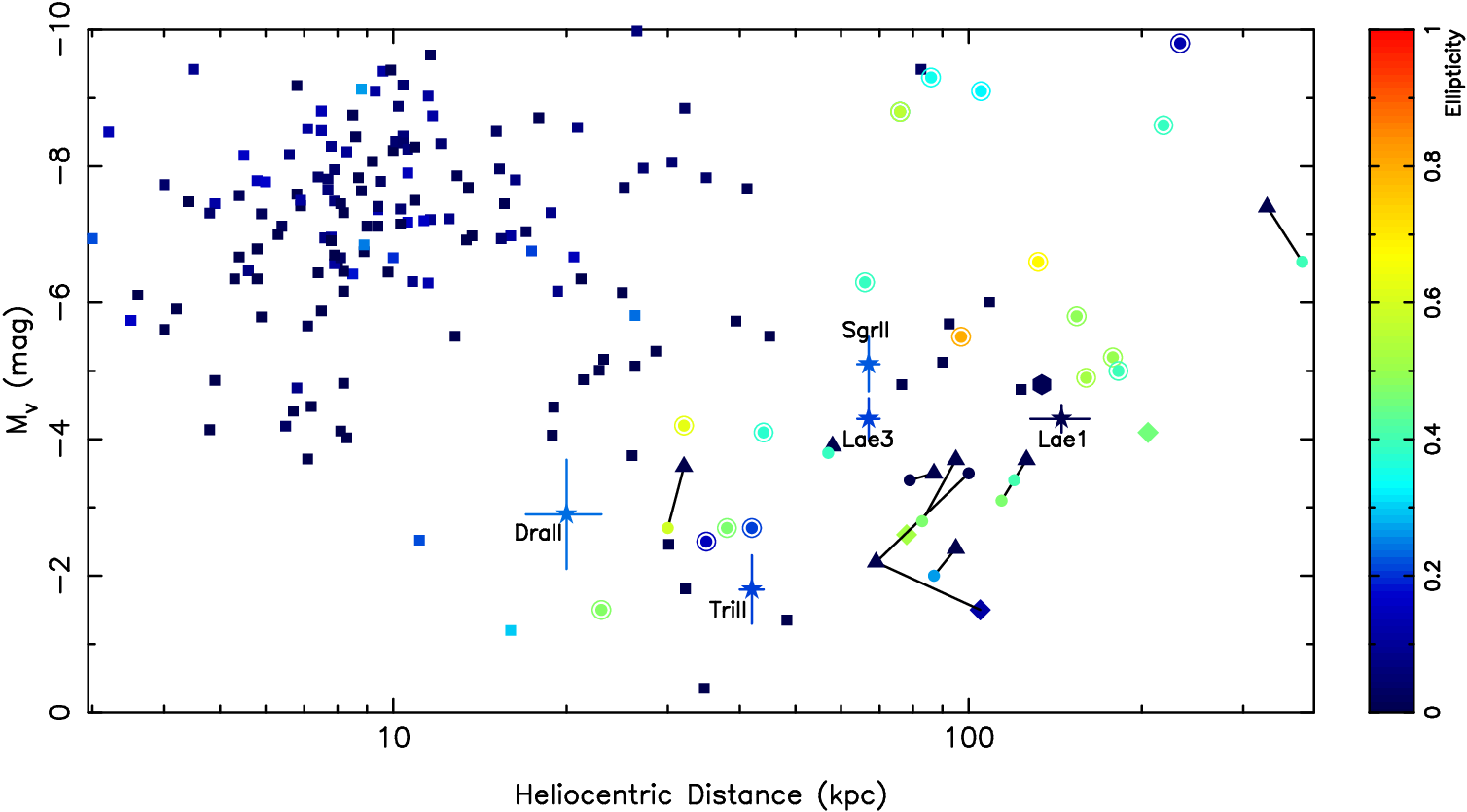}
  \caption{\label{rhMvDist}\emph{Top:} Distribution of MW satellites in the size-luminosity plane, color-coded by their ellipticity. Squares represent GCs from the \citet{harris10} catalog, supplemented by the more recent discoveries of Segue~3 \citep{belokurov10}, Mu\~{n}oz~1 \citep{munoz12b}, and Balbinot~1 \citep{balbinot13}. Milky Way confirmed dwarf galaxies are shown as circled dots, with their properties taken from \citet{mcconnachie12}. The co-discoveries by \citet{bechtol15} and \citet{koposov15} are shown with triangles and filled circles respectively, with the co-discoveries linked to each other by a black solid line reflecting the two groups' different measurements. The \citet{kim15a}, \citet{kim15},and \citet{kim15b} satellites are shown with diamonds. Hydra~II, discovered in SMASH is shown by a hexagon. Finally, the five PS1 discoveries (Lae~1, Tri~II, Sgr~II, Dra~II, and Lae~3) are shown as stars. \emph{Bottom:} The same for the size-Heliocentric distance plane.}
\end{figure*}

\section{Discussion}
Figure~\ref{rhMvDist} displays the properties of the three new discoveries in the context of the other MW satellites (GCs or DGs). The top panel shows the size-luminosity plane while the bottom panel focuses on the distance-luminosity plane. These parameters can already be a first indicator as to the nature of these objects, which we proceed to discuss here as well as the possible stream associations these objects may have.

\begin{figure}
\includegraphics[width=.58\textwidth,angle=270]{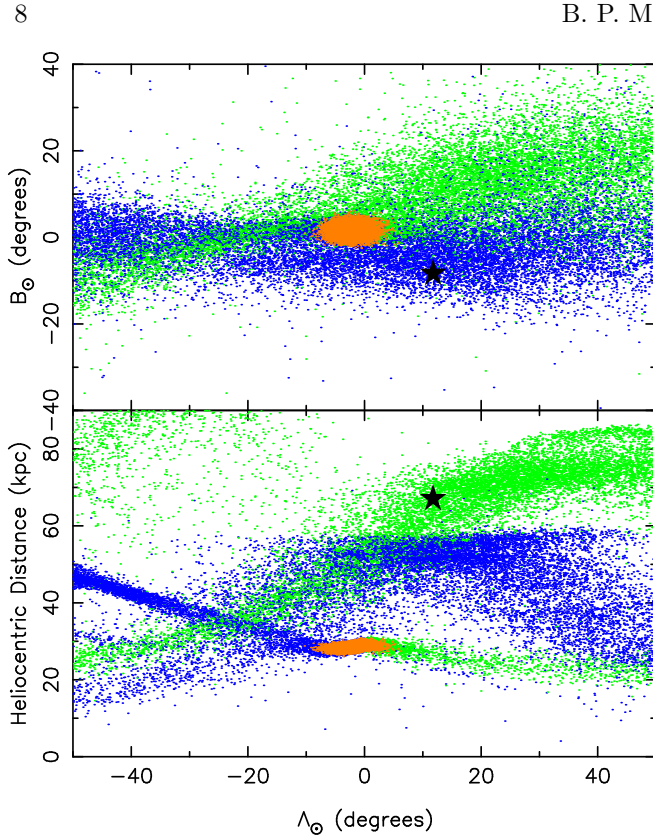}
\caption{\label{Sgr_Stream} \emph{Top:} Particles of the \citet{law10a} N-body model of the Sgr stream, projected on the Sgr dSph co-ordinate system \citep{majewski03}. Particles of the leading/trailing arm of the Sgr stream are shown in blue/green, whereas the body of the Sgr dSph is shown in orange. The position of Sgr~II is represented by the black star. \emph{Bottom:} The same for the Heliocentric distance vs. Sgr dSph longitude plane. Sgr~II clearly overlaps with the trailing arm.}
\end{figure} 

\begin{figure}
\includegraphics[width=.38\textwidth,angle=270]{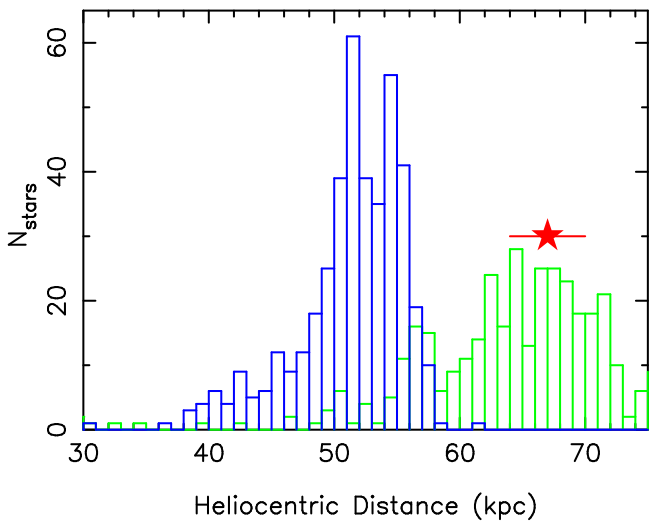}
\caption{\label{Sgr_Histo} Heliocentric distance histogram of all particles in the \citet{law10a} model within 1 degree of Sgr~II's longitudinal position (no constraint was applied on the latitude). As in Figure~\ref{Sgr_Stream}, blue and green dots represent particles from the leading and trailing arm, respectively. The distance to Sgr~II and its uncertainty are represented by the red star and the error bar and show that Sgr~II is perfectly compatible with belonging to the trailing arm of the Sgr stream.} 
\end{figure} 

\subsection{Sagittarius~II}
Sgr~II occupies an interesting place in the $r_{h}$ vs. $M_V$ plane as it lies in the very middle of the ``valley of ambiguity\footnote{The region in $r_{h}$ vs. $M_{V}$ space that straggles the `classical' boundaries between DGs and GCs.}'' highlighted by \citet{gilmore07}. Although other MW satellites are known with similar absolute magnitudes, Sgr~II is smaller than Coma Berenices ($r_h=74\pm4\pc$; \citealt{munoz10}, Pisces~II ($r_h=58\pm10\pc$; \citealt{sand12}), Hydra~II ($r_h=68\pm11\pc$; \citealt{martin15}), or the even larger Leo~IV and Leo~V ($r_h=205\pm36$ and $133\pm31\pc$; \citealt{dejong10a}), or Ursa Major~I ($r_h = 318\pm45\pc$; \citealt{martin08b}). On the other hand, Sgr~II remains larger than the largest GC, Pal~14 ($r_h\sim25\pc$; \citealt{hilker06}), or the recently discovered Laevens~1/Crater system ($r_h = 20\pm2\pc$), recently confirmed to be a GC \citep{kirby15}. It should however be noted that, recently, M31 satellite assumed to be GC have been discovered with similar sizes \citep{huxor14}, although the nature of some of these systems is also ambiguous \citep{mackey13}. The CMD of Figure~1 shows that the satellite hosts a clear blue HB, which is not a common feature of outer halo GCs that tend to favor red HBs (see, e.g., Figure~1 of \citealt{laevens14}). Ultimately, spectroscopic follow-up and a measure of the satellite's velocity dispersion is necessary to fully confirm the nature of this satellite and whether it is dark-matter dominated.

The location of Sgr~II, $\sim15\deg$ from Sgr dSph and in the expected plane of the Sgr stellar stream is particularly interesting as it could point to an association. The fact that it lies 40--45 $\kpc$ behind Sgr dSph rules out a direct connection between the two satellites but a comparison with the \citet{law10a} N-body model for the Sgr stream (Figure~\ref{Sgr_Stream}) reveals that Sgr~II is located at the expected distance of model particles from the trailing arm of the Sgr stream stripped out of their host more than $3\Gyr$ ago. It is therefore likely that Sgr~II was brought into the MW halo along with this part of the Sgr stream that has so far eluded detection, in a similar fashion to numerous other MW halos GCs \citep[e.g.][]{law10b}. The fact that the sky location of Sgr~II is slightly offset from this section of the model's trailing arm is not necessarily surprising since a former Sgr dSph satellite is not expected to be as concentrated on the sky as its former stars in the model. In addition, the location of these older wraps of the Sgr stream is very poorly constrained in the model. In fact, the discovery of Sgr~II and its association with the Sgr stream could add valuable constraints on the modeling of the Sgr stream once confirmed through radial velocities.

longitu
\subsection{Draco~II}
Draco~II also has an ambiguous nature, although it is here driven mainly by the large uncertainties on its structural parameters and distance, stemming from the faint nature of the object in the PS1 data. With the current photometry, the satellite appears to share the properties of Kim~2 or Eridanus~III, believed to be GCs \citep{kim15,bechtol15,koposov15}. On the other hand, its uncertain properties are also completely compatible with those of Wil~1, favored to be a DG \citep{willman11}. Here as well, deep photometry and/or spectroscopy are necessary to classify this system.

Given the common connection between faint stellar systems and stellar streams, we investigate possible associations of the satellite to known MW halo streams. The closest stream to Dra~II is the GD-1 stream \citep{grillmair06b}. Placing the new satellite onto the stream coordinate system $(\phi_{1},\phi_{2})$\footnote{This coordinate system is a rotated spherical one, aligned with the stream's coordinates. $\phi_{1}$ and $\phi_{2}$ represent longitude and latitude respectively.} defined by \citet{koposov10}, we find that it is located at $\phi_{1}\sim17.1^{\circ}$ and $\phi_{2}\sim3.8^{\circ}$. Though \citet{koposov10} do not have any measurements in this region (their measurements range from $\phi_1=-60.00^{\circ}$, $\phi_2=-0.64^{\circ}\pm0.15^{\circ}$ to $\phi_1=-2.00^{\circ}$, $\phi_2=-0.87^{\circ}\pm0.07^{\circ}$), the extrapolation of the orbit at the location of Dra~II yield $\phi_2\sim-2.7^{\circ}$, only 5--6$\deg$ away from the satellite. However, the extrapolated distance of the stream reaches only $\sim12\kpc$ there, to be compared with Dra~II's $\sim20\kpc$. Therefore, if the GD-1 stream does not significantly deviate from the \citet{koposov10} orbit, the current distance estimate for Dra~II  appears too high for a direct association.

\subsection{Laevens~3}
The small half-light radius of Lae~3 ($7\pm2\pc$) places it well within the regime of GCs. With a relatively young age ($\sim9\Gyr$) and stellar populations that are not very metal-poor ($\FeH\sim-1.9$), it would be natural to classify Lae~3 as a ``young outer halo'' GC found in the outer region of the MW halo \citep{mackey05}. However, some caveats should be noted: the isochrone fit relies on the photometry currently available, which suffers from crowding. The presence of an RR Lyra star could be at odds with a young halo scenario since its presence would point to a system that is at least 10~Gyrs old. We find no possible connection of this new system with known stellar streams in the MW halo.

\section{Conclusion}
In this paper, we have presented the discovery of three new faint Milky Way satellites, discovered in the photometric catalog of the PS1 3$\pi$ survey. The characterization of Lae~3 suggests that it is a GC, with properties similar to `young outer halo' GCs. The two other systems, Dra~II and Sgr~II, have an ambiguous classification. Dra~II contains mainly main sequence stars, as well as a handful of probable RGB stars. It is very faint but its structural parameters are uncertain enough to prevent a classification as an extended GC or a compact dwarf galaxy. It is located close to the orbital path of the GD-1 stream but its distance is in disagreement with the expectations of the stream's orbit ($\sim20$ vs. $\sim12 \kpc$) and appear to rule out an association. Finally, Sgr~II is located in a part of the size-luminosity plane that contains no other known system, either more extended than known MW GCs, or more compact than known MW DGs in its luminosity range. Independently of its nature, Sgr~II is particularly interesting as it lies at the expected location of the Sgr dSph stellar stream behind the bulge. In particular, the distance to the new satellite favors a connection with the currently undiscovered part of the trailing arm of the Sgr stream produced by stars stripped from the dwarf galaxy more than $3\Gyr$ ago. Ultimately, spectroscopic follow-up will be necessary to conclusively establish the nature of the last two satellites or confirm their connection with the GD-1 and Sgr stellar streams.

\acknowledgments
We thank Paolo Bianchini, Mark Norris, and Dougal Mackey for their thoughts and reflections. B.P.M.L. acknowledges funding through a 2012 Strasbourg IDEX (Initiative d'Excellence) grant, awarded by the University of Strasbourg. N.F.M. and B.P.M.L. gratefully acknowledges the CNRS for support through PICS project PICS06183. H.-W.R. and E.F.T acknowledge support by the DFG through the SFB 881 (A3). E.F.B. acknowledges support from NSF grant AST 1008342. 

The Pan-STARRS1 Surveys have been made possible through contributions of the Institute for Astronomy, the University of Hawaii, the Pan-STARRS Project Office, the Max-Planck Society and its participating institutes, the Max Planck Institute for Astronomy, Heidelberg and the Max Planck Institute for Extraterrestrial Physics, Garching, the Johns Hopkins University, Durham University, the University of Edinburgh, Queen's University Belfast, the Harvard-Smithsonian Center for Astrophysics, the Las Cumbres Observatory Global Telescope Network Incorporated, the National Central University of Taiwan, the Space Telescope Science Institute, the National Aeronautics and Space Administration under Grant No. NNX08AR22G issued through the Planetary Science Division of the NASA Science Mission Directorate, the National Science Foundation under Grant No. AST-1238877, the University of Maryland, and Eotvos Lorand University (ELTE). 

%\bibliography{Biblio_2}

\bibliographystyle{apj}

% Bibtex will create a .bbs file in the directory and before sending to the editor, I should replace the bibliography call by this file.

\end{document}